\documentclass[preprint,showpacs,preprintnumbers,amsmath,amssymb]{revtex4}

\usepackage{graphicx}
\usepackage{dcolumn}
\usepackage{bm}
\usepackage{longtable}

\begin{document}

\preprint{APS/123-QED}

\title{The thermonuclear rate for the $^{19}$F($\alpha$,p)$^{22}$Ne
reaction at stellar temperatures}

\author{C. Ugalde}
 \email{cugalde@unc.edu}
\affiliation{%
Department of Physics and The Joint Institute for Nuclear Astrophysics, University of Notre Dame, Notre Dame, Indiana 46556, USA\\
and\\
Department of Physics and Astronomy, University of North Carolina, Chapel Hill, North Carolina 27599, USA
and Triangle Universities Nuclear Laboratory, Box 90308, Durham, North Carolina 27708, USA
}%

\author{R. E. Azuma}
\affiliation{%
Department of Physics, University of Toronto, Toronto, Ontario
M55 1A7, Canada,
and The Joint Institute for Nuclear Astrophysics, Notre Dame, Indiana 46556, USA
}%

\author{A. Couture}
 \altaffiliation[Now at ]{Los Alamos Neutron Science Center, Los Alamos National Laboratory, Los Alamos, New Mexico 87544, USA} 
\author{J. G\"orres}
\author{H. Y. Lee}
 \altaffiliation[Now at ]{Argonne National Laboratory, Argonne, Illinois 60439, USA}
\author{E. Stech}
\author{E. Strandberg}
\author{W. Tan}
\author{M. Wiescher}
\affiliation{%
Department of Physics and The Joint
Institute for Nuclear Astrophysics,\\
University of Notre Dame, Notre Dame, IN 46556, USA
}%

\date{\today}

\begin{abstract}
The $^{19}$F($\alpha$,p)$^{22}$Ne reaction is considered to be one of
the main sources of fluorine depletion in AGB and Wolf-Rayet
stars. The reaction rate still retains large uncertainties due to
the lack of experimental studies available. In this work the
yields for both exit channels to the ground state and first
excited state of $^{22}$Ne have been measured and several previously
unobserved resonances have been found in the energy range
E$_{lab}$=792-1993 keV. The level parameters have been determined
through a detailed R-matrix analysis of the reaction data and a new
reaction rate is provided on the basis of the available
experimental information.
\end{abstract}

\pacs{Valid PACS appear here}
\maketitle

\section{Introduction}
Fluorine is by far the least abundant of the elements with atomic
mass between 11 and 32. While several nucleosynthesis scenarios
have been proposed for the production of fluorine, a unique site
for the origin of this element has not been identified yet.

Presently three different scenarios are being discussed for the
origin of fluorine. One possible process is the neutrino
dissociation of $^{20}$Ne in supernovae type II
\cite{Woosley:1988}. A second scenario is the pulsating He-burning
stage in AGB stars \cite{Busso:2006} and the third possibility is
the hydrostatic burning of helium in Wolf-Rayet stars
\cite{Goriely:1989}. It may be possible that all three sites
contribute to the formation of fluorine in the universe
\cite{Renda:2004}. So far, the only extra solar system
objects in the galaxy where fluorine has been observed are AGB
stars \cite{Jorrisen:1992} and post-AGB star configurations
\cite{Werner:2005}.

\begin{figure}
\includegraphics[width=8.5cm]{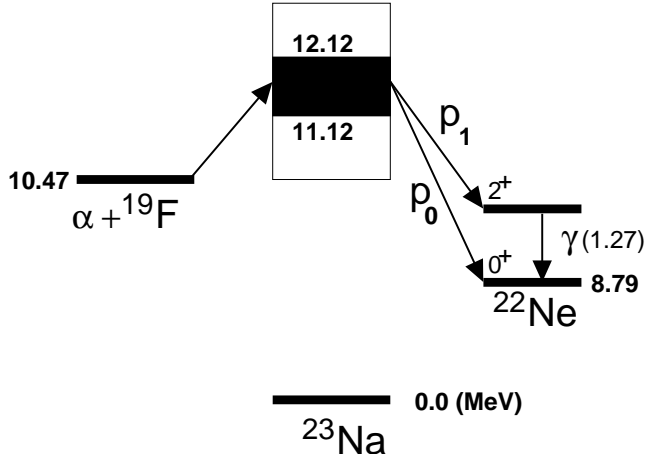}
\caption[Energy level scheme]
{Energy level scheme for the
$^{19}$F($\alpha$,p)$^{22}$Ne reaction. The entrance channel
($\alpha+^{19}$F) has a threshold of 10.47 MeV with respect to the
ground state of $^{23}$Na. The compound state can decay
either via the p$_0$ channel to the ground state of
$^{22}$Ne or by the p$_1$ channel to the first excited state of $^{22}$Ne.
The subsequent emission of a
1.27 MeV photon from the decay of the first excited state (2$^+$)
to the ground state (0$^+$) of $^{22}$Ne can also be observed. The 
dark rectangle above $^{23}$Na represents the energy region studied
in this work. White regions above and below it correspond to energies where 
the cross sections were extrapolated.}
\label{levels}
\end{figure}

Fluorine nucleosynthesis in AGB stars takes place in the 
hydrogen-helium intershell region where the $^{14}$N ashes from the preceding
CNO burning are converted to $^{18}$F by $\alpha$-particle 
captures (a similar reaction path is followed by
Wolf-Rayet stars). The unstable $^{18}$F isotope decays with a
half life of 109.8 minutes to $^{18}$O. Both a proton or an 
$\alpha$-particle can be captured by $^{18}$O. In the former case, 
$^{4}$He and $^{15}$N are being produced, while in the latter case
$^{22}$Ne is formed. This second possibility does not produce fluorine.
The alternative $^{18}$O(p,$\alpha$)$^{15}$N($\alpha,\gamma$)$^{19}$F capture
reaction is the main fluorine production link in this scenario.
The abundance of $^{15}$N depends sensitively on the hydrogen
abundance in the inter-shell region. Hydrogen has been depleted in
the preceding CNO burning but can be regenerated through the
$^{14}$N(n,p)$^{14}$C reaction, with the neutrons being produced by
the $^{13}$C($\alpha$,n)$^{16}$O reaction. Additional protons may be
mixed into the region when the convective envelope penetrates the
intershell region at the end of the third dredge up (TDU). 
$^{14}$C produced by this reaction provides a second link for
the production of $^{18}$O via $^{14}$C($\alpha,\gamma$)$^{18}$O.

Fluorine is very fragile and three reactions may cause rapid fluorine
destruction. Because of the high abundance of $^4$He in the
intershell region, the $^{19}$F($\alpha$,p)$^{22}$Ne is expected to be a dominant
depletion link. Another depletion process corresponds to the
$^{19}$F(n,$\gamma$)$^{20}$F reaction, caused by neutrons being produced by
the $^{13}$C($\alpha$,n)$^{16}$O or the $^{22}$Ne($\alpha$,n)$^{25}$Mg
reactions. If hydrogen is available in sufficient abundance,
fluorine may also be depleted through the very strong
$^{19}$F(p,$\alpha$)$^{16}$O reaction. All this, however, is strongly
correlated with the existing proton and $\alpha$-particle abundance 
at the fluorine synthesis site.

The reaction rate for $^{19}$F($\alpha$,p)$^{22}$Ne is highly
uncertain at helium burning temperatures. Even recent
nucleosynthesis simulations \cite{Palacios:2005} still rely on the
very simplified rate expression of Caughlan and Fowler (CF88) \cite{Caughlan:1988} based
on an optical model approximation for estimating the cross section
of compound nuclear reactions with overlapping resonances. This
approach was originally developed \cite{Fowler:1964} and employed
\cite{Wagoner:1969, Fowler:1975} for cases where no experimental
information was available. This reaction rate is in reasonable
agreement with more recent Hauser-Feshbach estimates assuming a
high density of unbound states in $^{23}$Na \cite{Thielemann:1986}. 
Other $^{19}$F nucleosynthesis simulations \cite{Lugaro:2004,Stancliffe:2005}
rely on the sparse experimental data available for
$^{19}$F($\alpha$,p)$^{22}$Ne for E$_{lab}$=1.3 MeV to
3.0 MeV \cite{Kuperus:1965}, estimating also possible low energy
resonant contributions from known $\alpha$-particle unbound states in
$^{23}$Na \cite{Endt:1990}. For example, in their approach they approximated the
reaction rate by only considering the resonance contributions of
the various states neglecting possible interference effects and
broad-resonance tail contributions. In this paper we describe a new
measurement of $^{19}$F($\alpha$,p)$^{22}$Ne at lower energies.
In our study we explored the reaction for
E$_{lab}$=792 to 1993 keV at different angles. 
The pronounced resonance structure was analyzed in
terms of the multi-channel R-matrix model using the recently
developed R-matrix code AZURE \cite{Azuma:2005}. Energy regions 
not measured here were also included in the computation
of the stellar reaction rate by using a combination of data 
available in the literature and our experimental results to 
extrapolate the reaction cross section. Possible low
energy resonances were taken into account sampling characteristic
$\alpha$-particle partial widths of the known $\alpha$-particle 
unbound states in $^{23}$Na, while deriving the corresponding 
proton partial width from the available elastic scattering data 
in the literature \cite{Keyworth:1968}. Interference terms were 
modelled with Monte Carlo simulations. In section 2
we describe the experimental setup and the experimental
procedure. The R-matrix analysis of the experimental data and 
the determination of the nuclear parameters of resonances is
described in section 3. Finally, the experimental and extrapolated
reaction rates (with error bars) are presented and discussed in 
section 4.

\section{Experimental Method}

The destruction of $^{19}$F by an $\alpha$-particle capture occurs 
mainly by a resonant reaction process through the $^{23}$Na compound nucleus
in an excitation range of high level density. The populated
resonant levels will decay by proton emission to the ground state
(p$_0$) or first excited state (p$_1$) of $^{22}$Ne. This study
included the direct measurement of the $^{19}$F($\alpha$,p)$^{22}$Ne
reaction by detecting both p$_0$ and p$_1$ protons (see \cite{Ugalde:2005b}
and \cite{Ugalde:PhDThesis}) as well as the additional measurement of the p$_1$ channel 
via the detection of the $\gamma$-ray transition \cite{Ugalde:2006} from the 
first excited state (2$^+$) to the ground state (0$^+$) of $^{22}$Ne 
(see figure \ref{levels}).

The experiment was performed at the Nuclear Science Laboratory of
the University of Notre Dame using the 3.5 MV KN Van de Graaff
accelerator. In a first run, the excitation curve of
$^{19}$F($\alpha$,p)$^{22}$Ne was investigated between E$_{lab}$=1224 keV and
1993 keV. For this experiment two Si 
surface barrier detectors with a depletion depth of 300 $\mu$m were mounted at forward
angles, while one 100 $\mu$m Si detector was positioned at a
backward angle. These thicknesses were sufficient for stopping the
reaction protons. The effective solid angle of the detectors at
each position configuration was determined using a mixed
$^{241}$Am and $^{148}$Gd $\alpha$-particle source with a 
known activity placed at the target position.

The energy range between E$_{lab}$=1629 keV to 1993 keV was mapped in 
5 keV energy steps with the detectors mounted at 30$^o$, 
90$^o$ and 130$^o$. This made it possible to use the two 
known resonances at E$_{lab}$=1.67 MeV and 1.89 MeV \cite{Kuperus:1965} 
as reference for calibrating and matching the reaction yield 
to the previous results. At lower energies, the detector position 
was changed to 40$^o$, 100$^o$ and 120$^o$ with respect to the 
beam direction. The excitation curve was mapped from E$_{lab}$=1220
to 1679 keV using the E$_{lab}$=1.67 MeV resonance as a reference. 
In total 483 proton spectra were acquired and for every energy, 
one elastic scattering $\alpha$-particle spectrum was taken. The 
stoichiometry of targets was constantly monitored with back-scattered 
$\alpha$-particles.

The $^{19}$F transmission targets were prepared by evaporating 10
$\mu$g/cm$^2$ of CaF$_2$ on 40 $\mu$g/cm$^2$ natural carbon substrates, 
mounted on aluminum frames. The target was placed with the
evaporated material facing the beam on a ladder attached to a
rotating rod at the center of the scattering chamber. The 
ladder held one target and a collimator that was used for centering the
beam. The targets deteriorated significantly under beam 
bombardment, so their stability was monitored by measuring
frequently the yield of the elastically scattered 
$\alpha$-particles at the E$_{lab}$=1.89 MeV resonance. Targets were 
constantly replaced with new and recently evaporated targets. 

For any two detectors with the same absolute efficiency the
relative count rate is independent of target stoichiometry and
beam intensity, as expressed by
\begin{equation}
{N_1(E,\theta)\over N_2(E,\theta)} = \biggl[{d\Omega_1 \over
d\Omega_2}\biggl]_{lab \over cm} \biggl({d\sigma(E,\theta) \over
d\Omega}\biggl)_2^{-1} \biggl({d\sigma(E,\theta) \over
d\Omega}\biggl)_1.
\end{equation}
$N_1(E,\theta)$ and $N_2(E,\theta)$ corresponds to the number of
events in detectors 1 and 2, respectively, $\bigl[$$d\Omega_1
\over d\Omega_2$$\bigl]$$_{lab/cm}$ is the solid angle correcting
for center of mass to the laboratory system, and
$d\sigma(E,\theta) \over d\Omega$ are the differential cross
sections measured at detectors 1 and 2, respectively. The
differential cross section of the $^{19}$F($\alpha$,p)$^{22}$Ne
reaction was determined relative to the differential cross section
for elastic scattering measured at 160$^o$. It has been shown by
Huang-sheng {\it et al.} \cite{Huan-sheng:1994} and by Cseh {\it et al.}
\cite{Cseh:1984} that below E$_{lab}$=2.5 MeV the elastic scattering of
$^4$He on $^{19}$F follows the Rutherford law,
\begin{equation}
{d\sigma(E,\theta) \over d\Omega}_{el}= {d\sigma(E,\theta) \over
d\Omega}_{Ruth}= \biggl({{Z_1 Z_2 e^2}\over{4E}}\biggl)^2
\sin^{-4}{\theta \over 2}.
\end{equation}
Z$_1$ and Z$_2$ are the atomic numbers of projectile and target,
respectively, $e$ is the proton electric charge and e$^2$=1440 keV fm, 
E is the relative energy of target and projectile in the
center of mass system, and $\theta$ is the center of mass angle at
which the elastically scattered particles are observed. Within the
small thickness of the target (27$\pm$5 keV) the stopping cross section
$\epsilon$ is assumed to be a constant. The variation of the
elastic differential cross section across the target thickness is
also very small. Therefore the elastic yield can be expressed as
\begin{equation}
Y_{elas}=\biggl({d\sigma(E,\theta) \over d\Omega}\biggl)_{Ruth}
{\Delta \over \epsilon}.
\label{elastic_yield}
\end{equation}

Subsequently the target integrated proton yield Y$_p$ can be
expressed relative to the elastic scattering cross sections as
\begin{widetext}
\begin{equation}
Y_p= {\Delta \over \epsilon} \biggl({d\sigma(E,\theta) \over
d\Omega}  \biggl)_{prot}= \biggl({N_{prot}(E,\theta)\over
N_{elas}(E,\theta)}  \biggl) \biggl({d\Omega_{prot} \over
d\Omega_{elas}} \biggl)_{lab/cm} {\Delta \over \epsilon}
\biggl({{Z_1 Z_2 e^2}\over{4E}}\biggl)^2 \sin^{-4}{\theta \over
2}.
\end{equation}
\end{widetext}

The energy dependence of the stopping power $dE/dx$ of $^4$He is
well known for both calcium and fluorine \cite{Ziegler:1985}. The
stopping power for fluorine was fitted to a quadratic polynomial
function given by 
\begin{equation}
\biggl({dE \over dx}\biggl)_F =\sum_{i=0}^{2}a_iE^i.
\end{equation}
A similar relation was determined for the calcium stopping power.
The partial stopping cross section $\epsilon$ for each of the
nuclear species in the target is described by:
\begin{equation}
\epsilon= {1 \over n}\biggl({dE \over dx}\biggl),
\end{equation}
where
\begin{equation}
n=\nu \rho N_A / A
\end{equation}
with $\nu$ the number of atoms per molecule, $\rho$ the density of
the target (again assuming the evaporated material has the same
density as the powder used before target preparation), 
$N_A$ the Avogadro number, and $A$
the mass number. The total stopping cross section of the calcium
fluoride target depends critically on the target stoichiometry
\begin{equation}
\epsilon=\epsilon_F + {N_{Ca} \over N_{F}} \epsilon_{Ca}.
\end{equation}
The ratio ${N_{Ca} \over N_{F}}$ measured in the evaporated target
layer does not necessarily reflect the stoichiometry of the
material before being evaporated. It was reported in previous work
\cite{Lorenz_Wirzba:Thesis} that evaporated CaF$_2$ shows a
stoichiometric calcium to fluorine ratio of 1:1. The targets used
in the present experiment were tested using well known resonances
in the $^{19}$F(p,$\alpha\gamma$)$^{16}$O reaction. The results indicated
that the stoichiometry of the evaporated material is the same as
that of the CaF$_2$ powder \cite{Couture:2005}.

In a second set of experiments we measured the proton
yield at lower energies. The experimental setup was designed
to perform the measurements with higher beam currents and larger
detector solid angles to compensate for the drop in reaction cross
section.

The target chamber for this set of experiments allowed 
mounting of the target at two
different angles with respect to the beam: at 45$^o$ and
90$^o$. With the first option we measured scattering
angles below 90$^o$, while with the second other angles were
measured at a smaller effective target thickness.
We tested several $^{19}$F-implanted targets for stability.  
Substrates tested were Ta, Ni, Cr, Al, Fe, and Mo. The best 
stability against beam deterioration was obtained for the Fe 
substrate. Electron suppression was supplied through an aluminum 
plate at -400 volts, mounted 5 mm in front of the target. Carbon
buildup on the target was minimized with a liquid nitrogen-cooled
copper plate. The target itself was water cooled from the back and
electrically isolated from the scattering chamber. Beam current
was directly measured at the target holder.

\begin{figure}
\includegraphics[width=8.5cm]{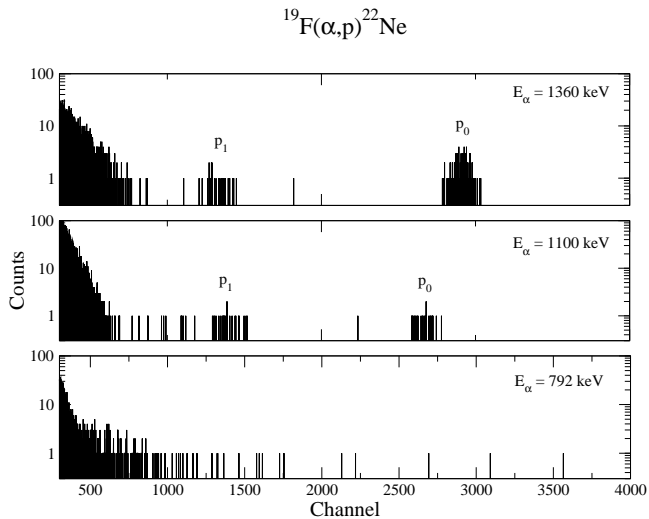}
\caption[Typical proton spectra] {Proton spectra for
$^{19}$F($\alpha$,p)$^{22}$Ne at 135$^o$ for three beam energies
(label at the right upper corner of each box). The upper spectrum
shows both proton groups at the reference resonance. Both peaks
appear clean of background and are easy to identify. The spectrum
at E$_{lab}$=1100 keV shows two groups of protons, still well isolated from
the background. Finally, a spectrum at E$_{lab}$=792 keV, where no proton
groups were positively identified is shown. The integrated charges
are (in $\mu$C) 1981, 180647, and 519252, respectively.}
\label{spectra}
\end{figure}

Two Si detectors were mounted on the rotating plate with aluminum
holders. Collimators were placed in front of the detectors and 
pin hole-free Al foils were used to stop the elastically scattered 
$\alpha$-particles, while allowing the protons to reach the surface
of the detectors. Both detector holders and the rotating plate 
were electrically isolated from the rest of the chamber. The 
detectors had an effective detection area of 450 mm$^2$. The 
solid angles for both detectors were determined using
a mixed $^{241}$Am+$^{148}$Gd $\alpha$-particle source of known 
activity mounted at the target position. We measured them to 
be 0.130$\pm$0.026 and 0.133$\pm$0.027 steradians, respectively.

With the chamber at the perpendicular target position and both
detectors at 135$^o$, a total of 540 spectra were acquired for
laboratory energies between 792 and 1380 keV. Typical 
spectra are shown in figure \ref{spectra}; the last spectrum 
taken (E$_{lab}$=792 keV) does not show identifiable proton groups. 
Subsequently, the chamber was reoriented to the 45$^o$ 
target position. The detectors were mounted at 150$^o$ and 
120$^o$ with respect to the beam direction and 178 spectra 
were taken. Finally, the detectors were mounted at 75$^o$ 
and 105$^o$ with respect to the beam direction and 69 spectra 
were measured.

In this experimental configuration the reaction yield could not be
normalized to elastic scattering because of the thick target
backing material. The yield was therefore measured relative to the
accumulated charge on the target during each run. The reaction
yield Y$_p$($\theta$) for the detector mounted at angle $\theta$ is
derived from the number of registered proton events N$_p$($\theta$)
in the detector is

\begin{equation}
Y_p(\theta)=\frac{N_p(\theta)}{N_{\alpha}\cdot\epsilon _p\cdot
d\Omega _p},
\end{equation}
where N$_{\alpha}$ is the accumulated charge, $\epsilon _p$ is the
absolute detection efficiency (assumed to be 1 for charged
particles and Si detectors at very low count rates), and
$d\Omega _p$ is the solid angle subtended by the detector.

The differential cross section can be directly derived from the
reaction yield normalized to the yield of the E$_{lab}$=1.37 MeV resonance
as measured in the first experiment. This depends critically on the
stability of the fluorine content in the target material. Since
the amount of fluorine decreases with accumulated charge,
the reaction yield needs to be corrected accordingly. During the
experiment the yield of the E$_{lab}$=1.37 MeV resonance was monitored
frequently to correct for target degradation. The on-resonance
thick target yield Y$_{tt}$ as a function of accumulated charge
N$_{\alpha}$ can be expressed by the linear relation:
\begin{equation}
Y_{tt} = a + bN_{\alpha},
\end{equation}
with a and b as constants. The measured yield of the observed
protons Y$_p$(E) was corrected for target degradation in terms of
the accumulated charge to
\begin{equation}
Y'_p(E) = \frac{a}{a+bN_{\alpha}}\cdot Y_p(E).
\end{equation}

Data obtained from all the described experiments consist of 20
excitation functions, with eleven corresponding to
$^{19}$F($\alpha$,p$_0$)$^{22}$Ne and nine to
$^{19}$F($\alpha$,p$_1$)$^{22}$Ne. Ten angles were measured in different
energy regions. All add up to 1471 data points (See table
\ref{tbl:curves}) which were analyzed in terms of the R-matrix
theory.

\begin{table}
\caption{\label{tbl:curves}Yield curves measured in this work}
\begin{ruledtabular}
\begin{tabular}{lccccc}
curve & channel & angle & E$_{min}$ (keV) & E$_{max}$ (keV) & $\Delta$ (keV)\\
\hline
1  & $^{19}F(\alpha,p_0)^{22}Ne$ & 130  &  1641 & 1993  & 15   \\
2  & $^{19}F(\alpha,p_0)^{22}Ne$ &  90  &  1641 & 1993  & 15   \\
3  & $^{19}F(\alpha,p_0)^{22}Ne$ &  30  &  1641 & 1993  & 15   \\
4  & $^{19}F(\alpha,p_0)^{22}Ne$ & 120  &  1224 & 1679  & 15   \\
5  & $^{19}F(\alpha,p_0)^{22}Ne$ & 100  &  1224 & 1679  & 15   \\
6  & $^{19}F(\alpha,p_0)^{22}Ne$ &  40  &  1224 & 1679  & 15   \\
7  & $^{19}F(\alpha,p_0)^{22}Ne$ & 105  &  1027 & 1367  & 25   \\
8  & $^{19}F(\alpha,p_0)^{22}Ne$ & 120  &   929 & 1359  & 35   \\
9  & $^{19}F(\alpha,p_0)^{22}Ne$ & 135  &   792 & 1363  & 25   \\
10 & $^{19}F(\alpha,p_0)^{22}Ne$ & 150  &   929 & 1359  & 35   \\
11 & $^{19}F(\alpha,p_0)^{22}Ne$ &  75  &  1027 & 1367  & 35   \\
12 & $^{19}F(\alpha,p_1)^{22}Ne$ & 130  &  1629 & 1981  & 15   \\
13 & $^{19}F(\alpha,p_1)^{22}Ne$ &  90  &  1629 & 1981  & 15   \\
14 & $^{19}F(\alpha,p_1)^{22}Ne$ &  30  &  1629 & 1981  & 15   \\
15 & $^{19}F(\alpha,p_1)^{22}Ne$ & 120  &  1224 & 1679  & 15   \\
16 & $^{19}F(\alpha,p_1)^{22}Ne$ & 100  &  1224 & 1679  & 15   \\
17 & $^{19}F(\alpha,p_1)^{22}Ne$ &  40  &  1224 & 1679  & 15   \\
18 & $^{19}F(\alpha,p_1)^{22}Ne$ & 120  &   929 & 1359  & 35   \\
19 & $^{19}F(\alpha,p_1)^{22}Ne$ & 135  &   792 & 1363  & 25   \\
20 & $^{19}F(\alpha,p_1)^{22}Ne$ & 150  &   929 & 1359  & 35   \\
\end{tabular}
\end{ruledtabular}
\end{table}

\section{Multichannel R-matrix analysis}

For the analysis of the experimental data we used the A-matrix version of the
computer code AZURE \cite{Azuma:2005}. AZURE is
a multichannel and multilevel code that implements the A- and R-matrix
formalisms as presented by Lane and Thomas \cite{Lane:1958}. The code is 
capable of fitting experimental datasets by varying formal parameters 
(energies and width amplitudes) of compound-nucleus states. The integrated 
cross section can also be computed from angular distribution datasets. 
Error bars for both the parameters and cross section curves are treated 
with Monte Carlo techniques.

The input to the code consists of a set of initial nuclear
parameters; each level is characterized by one energy eigenvalue
and several formal reduced width amplitudes (one per channel per
level). The code identifies open reaction channels and from the input 
it assigns an independent width amplitude for each (s,l) 
combination allowed for the level. Theoretical differential 
cross section curves at different angles are computed and then compared 
to experimental yields after target integration corrections \cite{Ugalde:2006a}. The 
maximum likelihood estimator is then minimized by varying all 
the parameters simultaneously. Each time a local minimum is found, 
the integrated value of the cross section is computed.

Overlapping of resonances complicated the simultaneous fitting of 
the yield curves. The code was by itself not able to find a set 
of formal parameters that would reasonably describe the yield curves. 
For this reason, an initial parameter set had to be found without 
the help of the minimization routines. By trial 
and error, the choice of initial nuclear parameters was done by adjusting 
the energy for each of the levels so as to describe the position of 
resonances as close as possible. The height and width of the
resonances was, on the other hand, approximated by varying the
width amplitudes.

Interference patterns between the various resonances were
determined by iteratively probing the contribution to yield curves of
levels within groups of the same J$^{\pi}$. The sign of the width 
amplitudes was flipped one by one for all the (s,l) channels.
These steps were repeated iteratively several times until the theoretical 
curves resembled the dataset. The resulting set of parameters was then used 
as input to the R-matrix code coupled to the $\chi^2$ minimization routines. 
Every time a calculation was performed all 20 excitation curves were 
examined. The target thickness $\Delta$ used for each of the yield curves 
is shown in table \ref{tbl:curves}.

\begin{widetext}

\begin{figure}
\includegraphics[width=15cm]{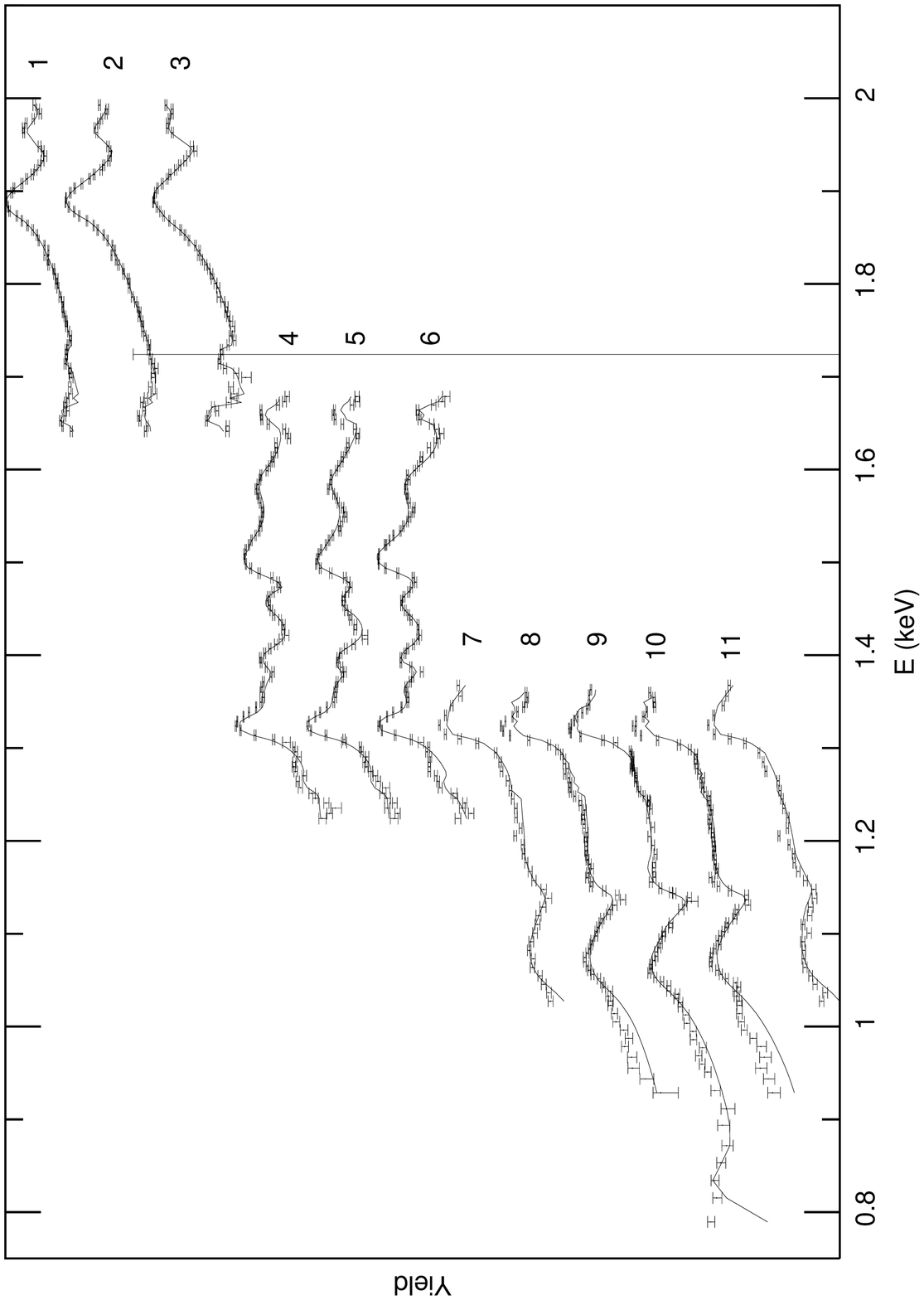}
\caption
{Experimental yield curves for the $^{19}$F($\alpha$,p$_0$)$^{22}$Ne channel
and the R-matrix fits (solid lines).
Different curves are offset by a factor of 100. Vertical axis units are
relative yields. The labels to the right of each curve are described in
table \ref{tbl:curves}.}
\label{p0}
\end{figure}

\begin{figure}
\includegraphics[width=15cm]{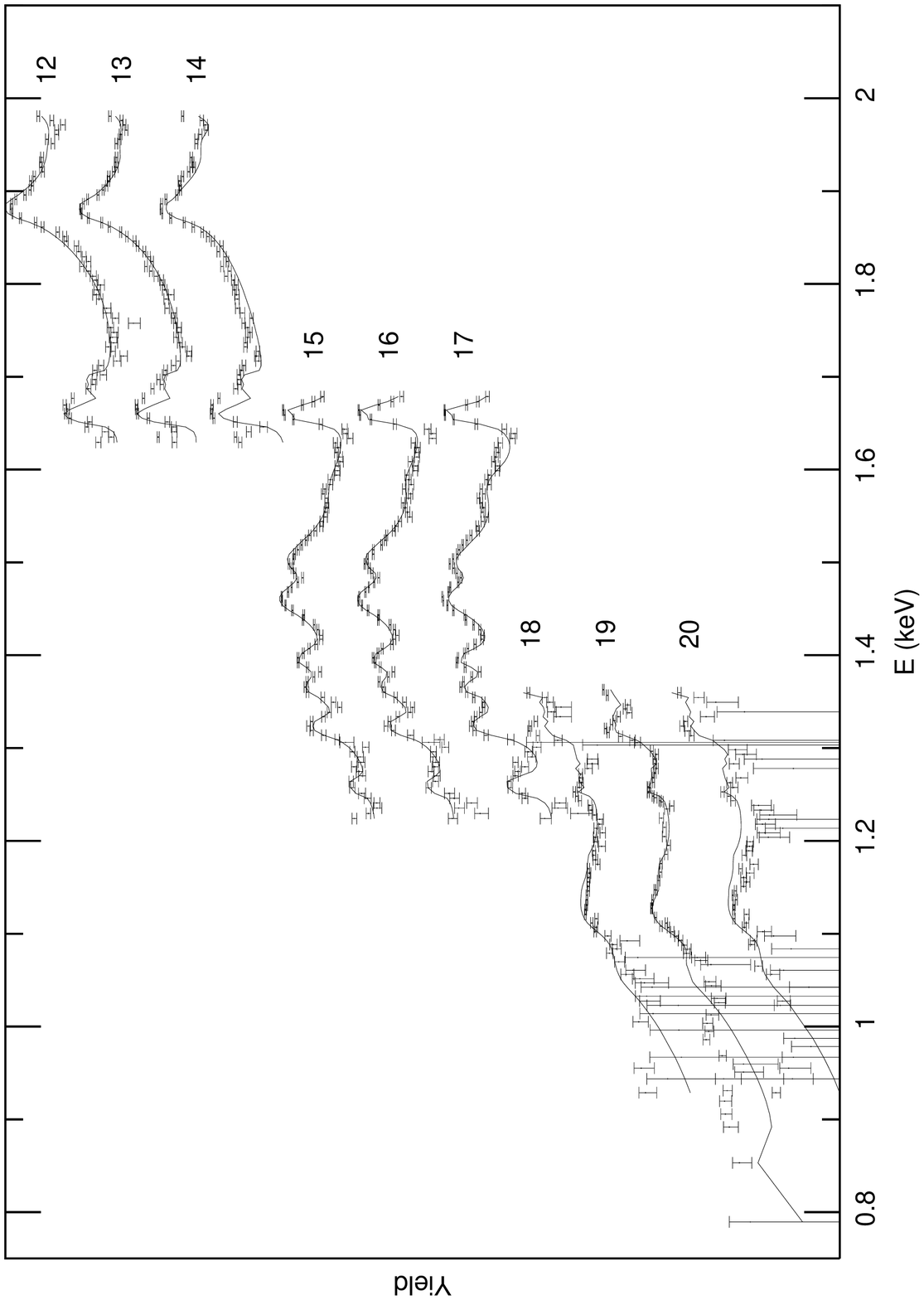}
\caption
{Experimental yield curves for the $^{19}$F($\alpha$,p$_1$)$^{22}$Ne channel and
the R-matrix fits (solid lines).
Different curves are offset by a factor of 100. Vertical axis units are
relative yields. The labels to the right of each curve are described in
table \ref{tbl:curves}.}
\label{p1}
\end{figure}

\end{widetext}

The resulting R-matrix fits to the experimental yield data $Y_p$
are shown in figures \ref{p0} and \ref{p1} for the ground state
and the excited state transitions, respectively. Both the 
energies E$_\lambda$ and reduced width amplitudes $\gamma_i$ 
were determined with a single channel radius 
(a$_c$=5.5 fm). The boundary condition was set to $S_c$, the shift function,
at the level in each J$^\pi$ group with the lowest energy. Background states were 
included for each of the J$^{\pi}$ groups. (The set of 201 R-matrix parameters 
and the experimental dataset can be obtained by contacting the author.)

\begin{figure}
\includegraphics[width=8.5cm]{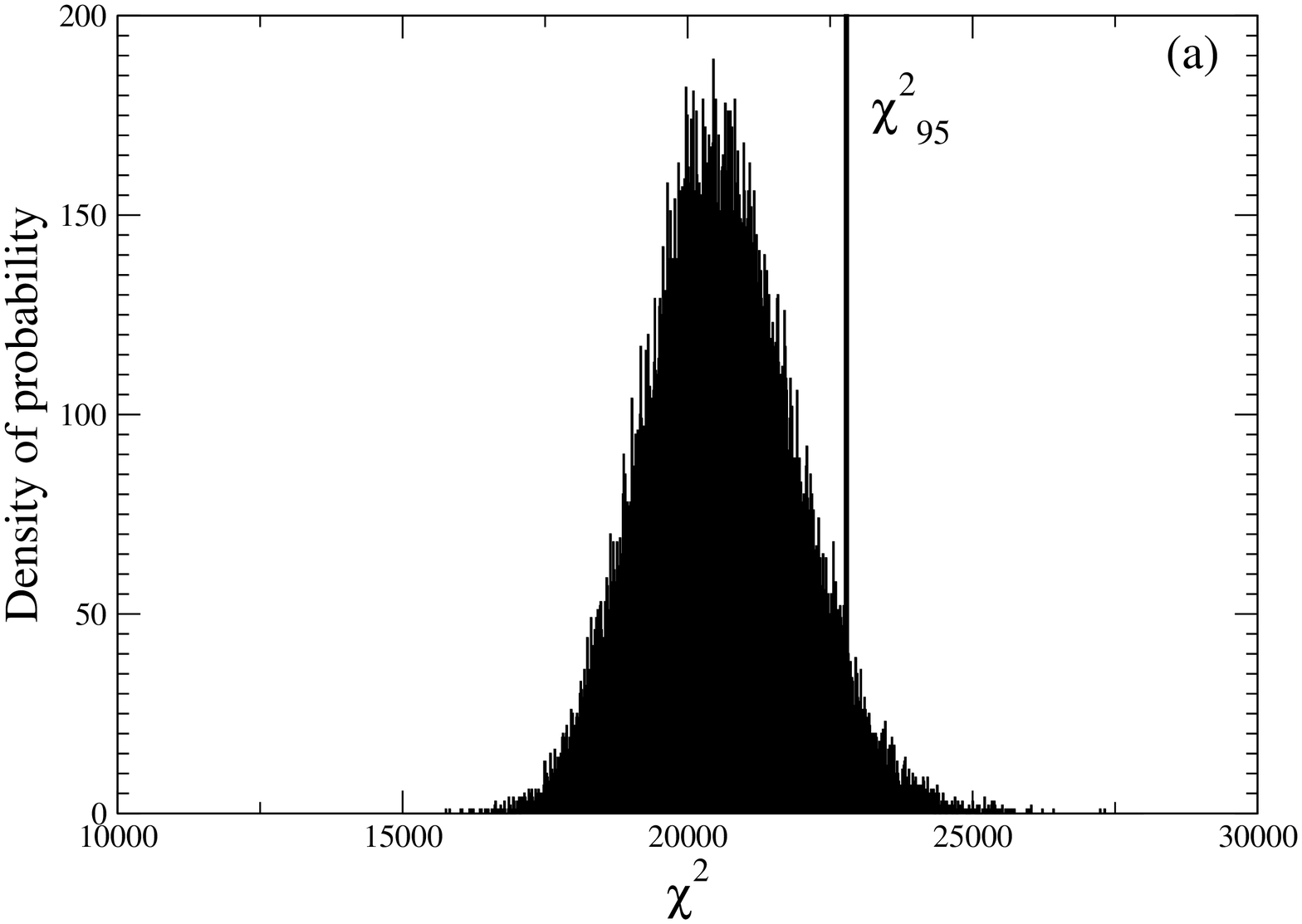}
\includegraphics[width=8.5cm]{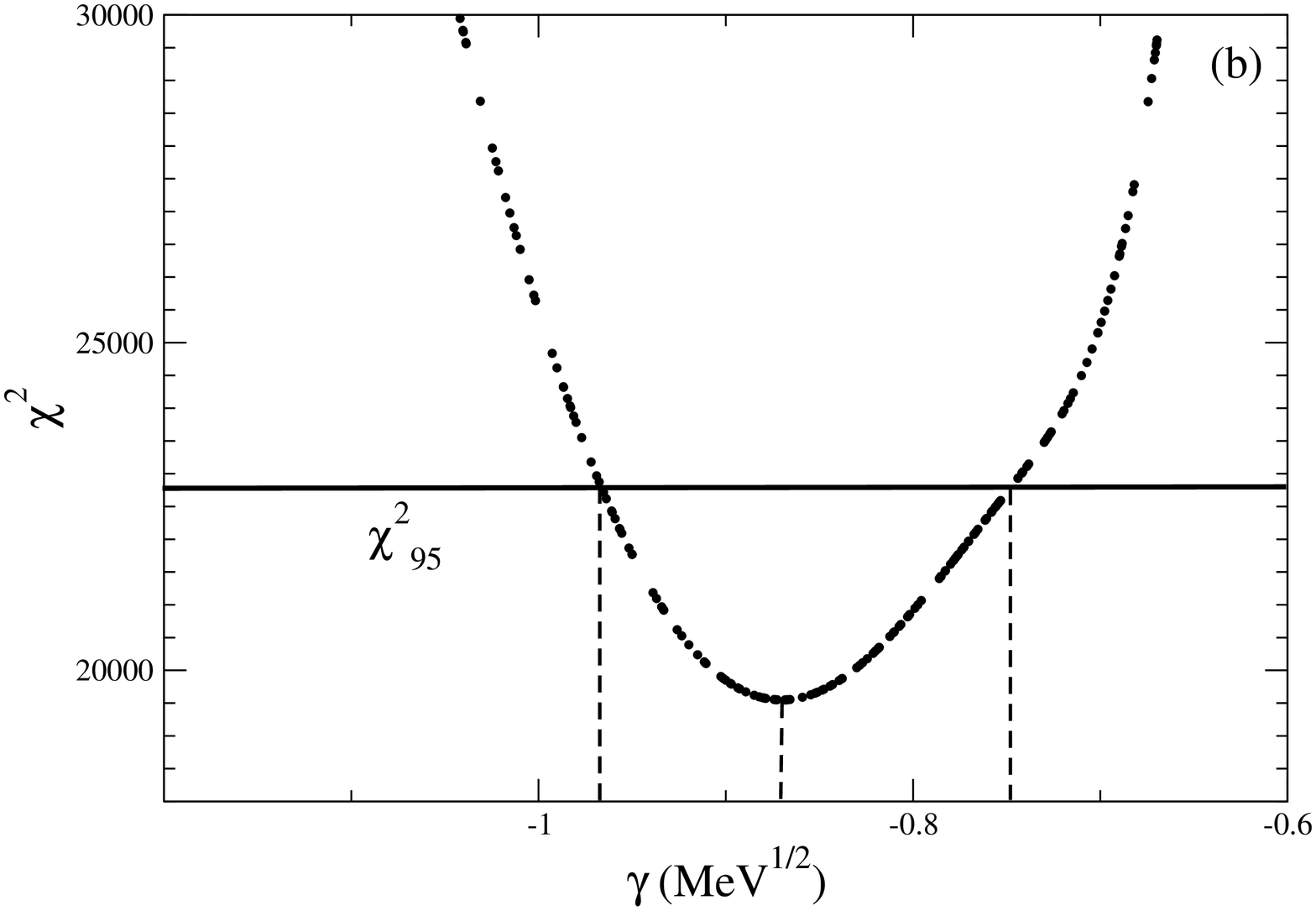}
\caption
{Calculation of error bars. (a) Bootstrap histogram used to compute the error bars
 with a 95\% confidence level. The horizontal axis 
corresponds to the raw $\chi^2$ value 
associated to the set of central-value parameters. Each bin contains intervals  
$\Delta \chi^2$=20. A total of 40,000 subsets were sampled from the experimental set
with a Monte Carlo method. Note that the curve does not correspond to a normal distribution.
However, the centroid corresponds to the raw value of $\chi ^2$ associated to the best R-matrix fit 
($\chi^2_{min}$=20,535). An area corresponding to 95\% of the total integral is
below $\chi^2_{95}$=22,774 (vertical line). (b) Sample $\chi ^2$ vs. parameter value ($\gamma$) curve.
The horizontal line represents $\chi^2_{95}$, while the dotted vertical lines are, from left to right,
the lower limit, central value, and upper limit of the parameter, respectively. An equivalent curve was
generated for each of the formal parameters.}
\label{bootstrap}
\end{figure}

Parameter error bars were determined with a bootstrap method \cite{Press:1992}
and correspond to a confidence interval of 95\% (see figure \ref{bootstrap}). 
From the set of 1,471 experimental data 
points we generated 40,000 subsets, with each subset containing 1,471 data 
points as well. The selection of points was performed with a Monte Carlo 
method and as a result, each set contains a random number of points repeated 
more than once. Using the set of formal parameters obtained with the R-matrix 
analysis described above, a $\chi ^2$ was computed for each of the subsets 
sampled by bootstrapping. The distribution of $\chi ^2$ values is shown in 
figure \ref{bootstrap}(a). The 0.95 cumulative value of the distribution corresponds 
to $\chi^2\equiv\chi^2_{95}$=22,774. Finally, with a Monte Carlo technique, we generated 
$\chi^2$ curves by varing each parameter independently while fixing all others.
The error bar corresponds to the largest parameter value such that 
$\chi^2$ $\leq$ $\chi^2_{95}$ (see figure \ref{bootstrap}(b)). 
Total cross sections for both channels 
were computed as well and are discussed in the next section and shown in 
figure \ref{theCrossSections}.

\section{The reaction rate}

The thermonuclear reaction rate has contributions from three
energy regimes: a) the region measured experimentally in this work,
spanning from E$_{lab}$=792 keV to 1993 keV, b) the region
below E$_{lab}$=792 keV (not measured), and c) the region above
E$_{lab}$=1993 keV (not measured here as well).

\begin{figure}
\includegraphics[width=9.5cm]{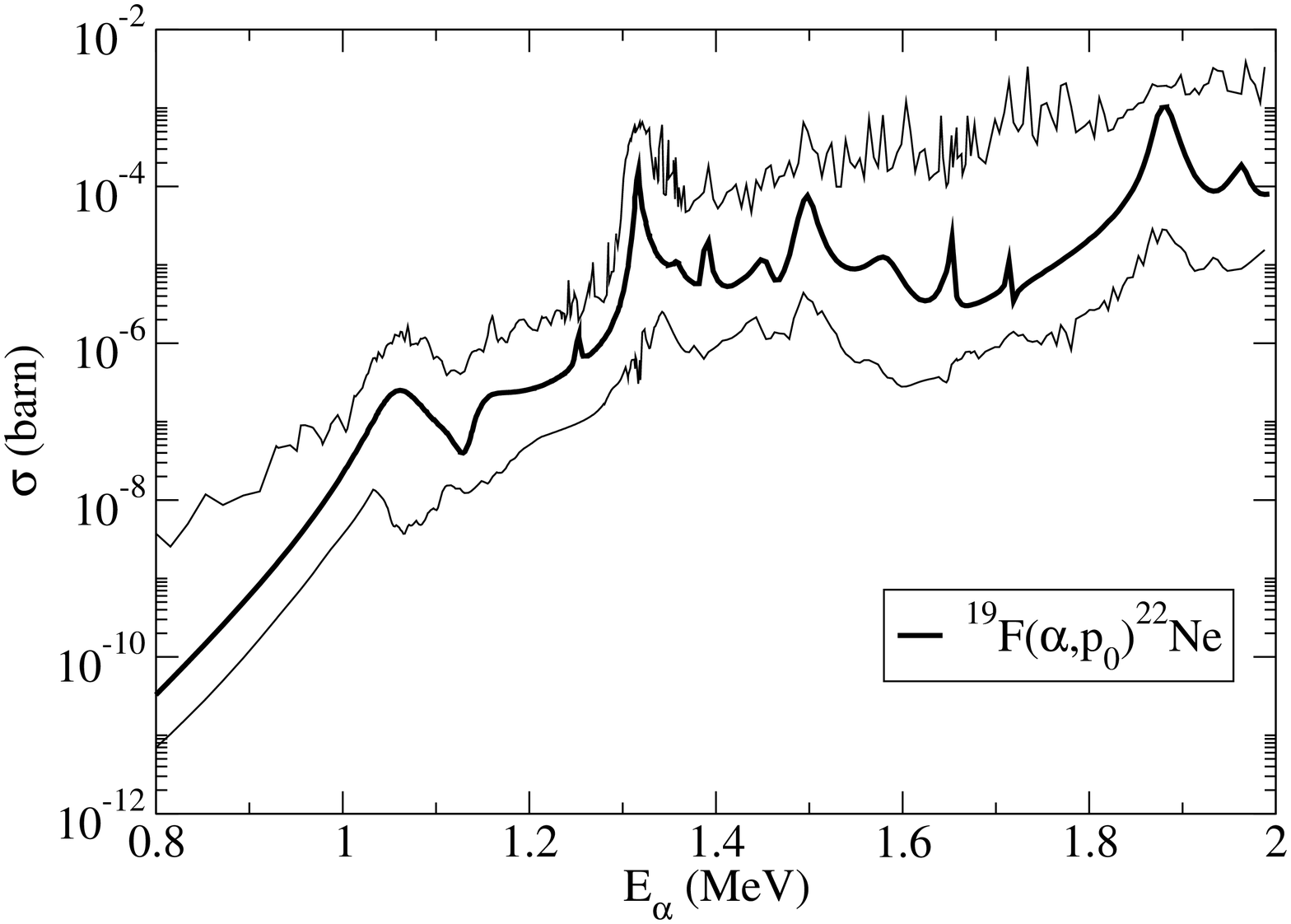}
\includegraphics[width=9.5cm]{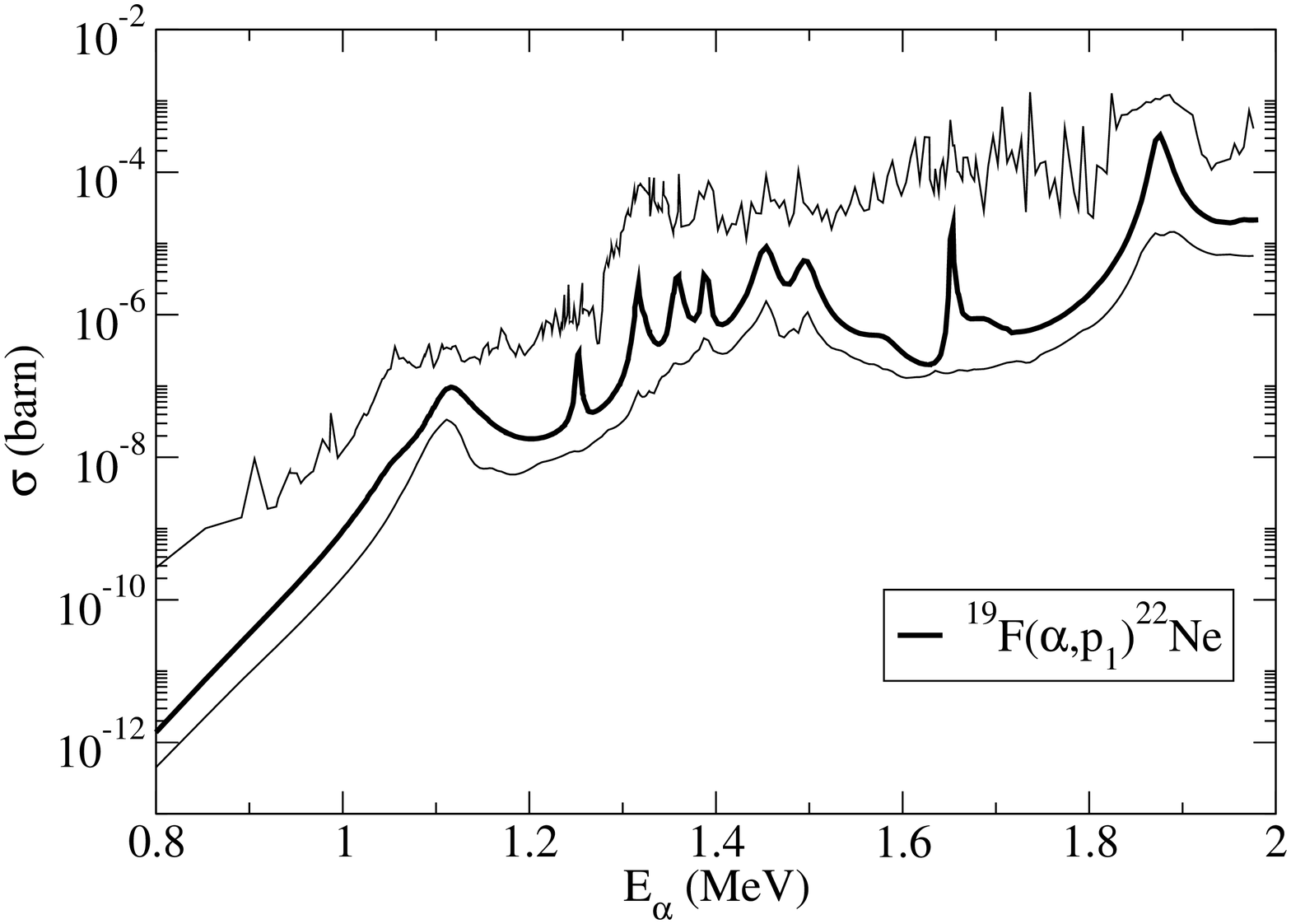}
\caption{R-matrix calculated total cross sections for $^{19}$F($\alpha$, p)$^{22}$Ne. Both 
$^{19}$F($\alpha$, p$_0$)$^{22}$Ne (upper panel) and $^{19}$F($\alpha$, p$_1$)$^{22}$Ne (lower panel) 
curves are shown with uncertainty
bands, as computed by sampling the 95\% confidence interval nuclear parameter space with a Monte 
Carlo method. The strong 
oscillations in the upper and lower limits of the bands are an artifact of the finite number 
of samples (10,000) taken with the Monte Carlo method.}
\label{theCrossSections}
\end{figure}

The contribution to the rate from our experimental
dataset was calculated by integrating numerically the total cross section
$\sigma(E)$ over the Maxwell-Boltzmann distribution of stellar gas
at temperature $T$
\begin{equation}
{N_A}\langle \sigma v
\rangle ={N_A}{ \biggl( {8\over{\pi \mu}} \biggl )^{1/2}
{(kT)^{-3/2}} {\int \sigma(E) E \exp \biggl({{-E}\over{kT}}
\biggl)dE}}. 
\label{rateBoltzmann}
\end{equation}
Here E is the energy of the particles in the center of mass system, N$_A$ is
Avogadro's number, k is Boltzmann's constant, $\mu$ the reduced mass, and T 
the temperature of the gas. The total cross section $\sigma$ was derived from 
the R-matrix calculation with the recommended values of formal parameters. 
Both p$_0$ and p$_1$ components are shown in figure 
\ref{theCrossSections}. Upper and lower limits of the total cross section and 
the reaction rate in this energy regime were computed by sampling the parameter space 
defined by the upper and lower values of the fitting parameters with a 
Monte Carlo technique. All parameters were varied simultaneously and a total  
of 10,000 parameter sets were produced. The reaction rate for each parameter 
set was calculated with equation \ref{rateBoltzmann} and all resulting rates 
were compared with each other. The highest (lowest) value obtained corresponds 
to the upper (lower) limit of the reaction rate.

The contribution to the reaction rate from resonances below E$_{lab}$=792 keV, 
which were not measured in the present experiment, was considered as well.
Previous studies through other reaction channels do indicate several unbound 
states in $^{23}$Na in this energy range near the $\alpha$-particle threshold 
\cite{Endt:1990}, which may contribute significantly to the 
$^{19}$F($\alpha$,p)$^{22}$Ne reaction. Most notably, detailed elastic proton 
scattering measurements were performed for this energy range in $^{23}$Na 
\cite{Keyworth:1968}, and provide important information necessary for estimating 
the contributions of these lower energy states to the reaction rate.
Resonances observed in the $^{22}$Ne(p,p)$^{22}$Ne and $^{22}$Ne(p,p')$^{22}$Ne  
channels were used to define the spins, parities, energies, and both proton p$_0$ and 
p$_1$ partial widths of contributing states. On the other hand, $\alpha$-particle 
partial widths $\Gamma_\alpha$ were obtained by adopting the experimentally known 
$\alpha$-particle reduced widths  $\gamma^2_\alpha$ determined in the high
energy range.

\begin{figure}
\includegraphics[width=8.5cm]{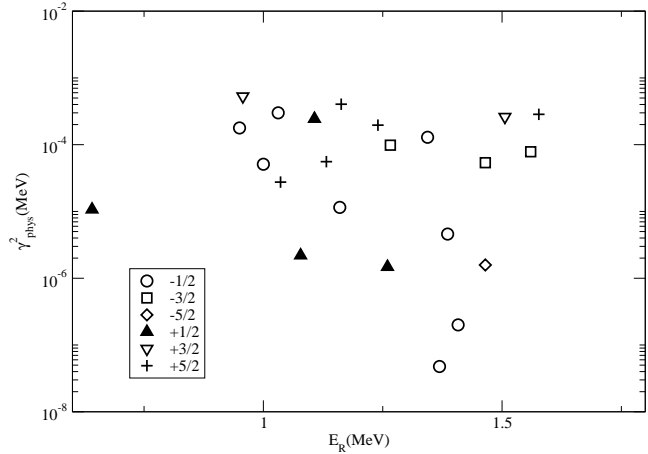}
\caption[Experimental reduced $\alpha$-particle widths for $^{19}$F($\alpha$,p)$^{22}$Ne]
{The experimentally determined $\gamma_\alpha ^2$ as a function of
center-of-mass energy, presented in sets of $J^\pi$. }
\label{theReducedWidths}
\end{figure}

Energies and reduced widths obtained with the R-matrix analysis
were used to calculate
the logaritmic average $<log (\gamma^2_\alpha)>$ for each
J$^{\pi}$ group (see figure \ref{theReducedWidths}).
The extrapolated reduced $\alpha$-particle width amplitude for
a state with parameters (J,$\pi$) was chosen to be
\begin{equation}
\gamma^2_\alpha(J,\pi) = 10^{<log (\gamma^2_\alpha)>}.
\end{equation}

Extrapolated upper (lower) limit values of the reduced $\alpha$-particle 
widths were set equal to the highest (lowest) $\gamma_\alpha ^2$
value determined from the experimental data of the corresponding
J$^\pi$ group. The set of extrapolated $\gamma^2_\alpha$ values,
together with the reduced proton widths $\gamma^2_p$
calculated from $^{22}$Ne(p,p)$^{22}$Ne and
$^{22}$Ne(p,p')$^{22}$Ne experiments by $\Gamma_p= 2P \gamma^2_p$
(where $P$ is the penetrability through the Coulomb barrier) is shown
in table \ref{tbl:extra}.

\begin{table}
\caption{\label{tbl:extra}Nuclear parameters for states
with E$_{lab}<$792 keV.}
\begin{ruledtabular}
\begin{tabular}{cccccc}
$J$ & $\pi$ & $E_{cm}$ $(MeV)$ & $\gamma^2_\alpha$ $(MeV)$ &
$\gamma^2_\alpha$ $(MeV)$ & $\gamma^2_\alpha$ $(MeV)$ \footnotemark[1] \\
      &      &        &     Recomm     &    Upper  &  Lower\\
\hline
1.5   &   1  & 0.010  &     3.72E-4     &   5.29E-4  &  2.61E-4\\
1.5   &   -1     & 0.031  & 8.89E-5     &   1.51E-4  &  5.37E-5\\
0.5   &   1  & 0.037  & 9.60E-6     &   2.44E-4  &  1.48E-6\\
2.5   &   1  & 0.049  & 1.04E-4     &   4.04E-4  &  2.75E-5\\
2.5   &   1  & 0.078  & 1.04E-4     &   4.04E-4  &  2.75E-5\\
1.5   &   -1     & 0.106  & 8.89E-5     &   1.51E-4  &  5.37E-5\\
1.5   &   1  & 0.147  & 3.72E-4     &   5.29E-4  &  2.61E-4\\
2.5   &   1  & 0.147  & 1.04E-4     &   4.04E-4  &  2.75E-5\\
1.5   &   -1     & 0.207  & 8.89E-5     &   1.51E-4  &  5.37E-5\\
1.5   &   -1     & 0.237  & 8.89E-5     &   1.51E-4  &  5.37E-5\\
1.5   &   1  & 0.354  & 3.72E-4     &   5.29E-4  &  2.61E-4\\
1.5   &   -1     & 0.355  & 8.89E-5     &   1.51E-4  &  5.37E-5\\
1.5   &   1  & 0.369  & 3.72E-4     &   5.29E-4  &  2.75E-5\\
2.5   &   1  & 0.369  & 1.04E-4     &   4.04E-4  &  2.61E-4\\
1.5   &   -1     & 0.405  & 8.89E-5     &   1.51E-4  &  5.37E-5\\
0.5   &   -1     & 0.438  & 1.07E-5     &   3.00E-4  &  4.76E-8\\
2.5   &   1  & 0.438  & 1.04E-4     &   4.04E-4  &  2.75E-5\\
0.5   &   1  & 0.448  & 9.60E-6     &   2.44E-4  &  1.48E-6\\
1.5   &   1  & 0.461  & 3.72E-4     &   5.29E-4  &  2.61E-4\\
0.5   &   1  & 0.481  & 9.60E-6     &   2.44E-4  &  1.48E-6\\
2.5    &  1  & 0.503  & 1.04E-4     &   4.04E-4  &  2.75E-5\\
1.5    &  1  & 0.503  & 3.72E-4     &   5.29E-4  &  2.61E-4\\
1.5    &  1  & 0.506  & 3.72E-4     &   5.29E-4  &  2.61E-4\\
1.5    &  -1     & 0.511  & 8.89E-5     &   1.51E-4  &  5.37E-5\\
0.5    &  1  & 0.524  & 9.60E-6     &   2.44E-4  &  1.48E-6\\
1.5    &  1  & 0.525  & 3.72E-4     &   5.29E-4  &  2.61E-4\\
0.5    &  1  & 0.569  & 9.60E-6     &   2.44E-4  &  1.48E-6\\
0.5    &  -1     & 0.618  & 1.07E-5     &   3.00E-4  &  4.76E-8\\
2.5    &  1  & 0.640  & 1.04E-4     &   4.04E-4  &  2.75E-5\\
1.5    &  1  & 0.642  & 3.72E-4     &   5.29E-4  &  2.61E-4\\
\end{tabular}
\footnotetext[1]{Center-of-mass energies, spins, and parities are
from \cite{Keyworth:1968}. The reduced $\alpha$-particle widths were obtained by
extrapolating the values measured in this work.}
\end{ruledtabular}
\end{table}

\begin{figure}
\includegraphics[width=8.5cm]{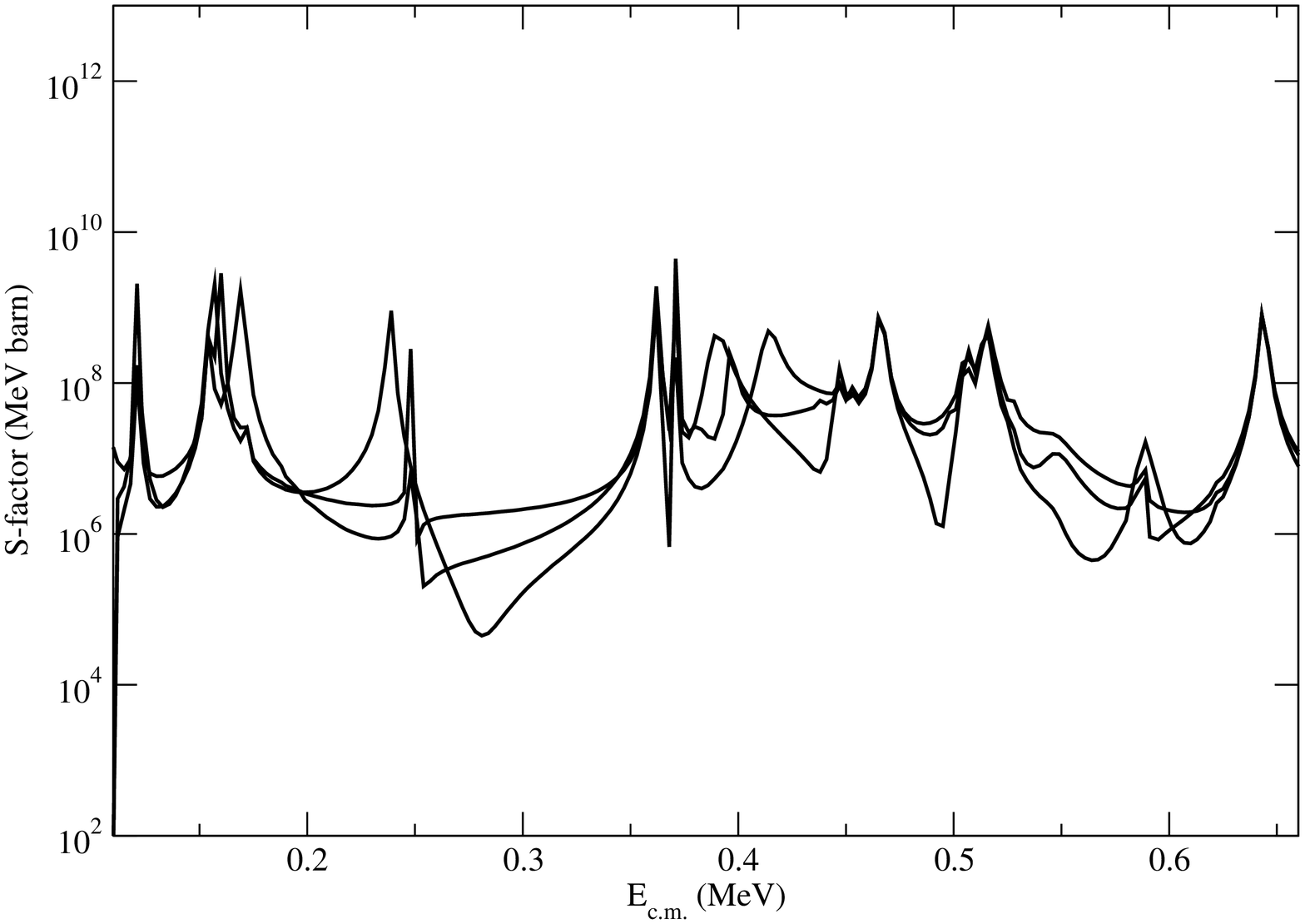}
\caption[Monte Carlo simulated low energy cross section]
{Monte Carlo simulated low energy S-factor for $^{19}$F($\alpha,p$)$^{22}$Ne
for three different sample resonance interference assumptions.}
\label{extraS}
\end{figure}

Based on these parameters, the resonant contribution to the 
cross section was calculated using the R-matrix formalism. 
Nevertheless, the interference pattern between resonances 
can not be predicted in this scheme. Therefore, we simulated its 
effect with a Monte Carlo sampling of possible width amplitude 
sign combinations. Figure \ref{extraS} shows the resulting 
S-factor for three sample assumptions for interference between 
resonances. Upper and lower values of the reaction rate
were calculated with equation \ref{rateBoltzmann} from 
different combinations of signs and widths 
sampled within the parameter space, as before. The recommended 
value corresponds to the logaritmic average of the upper and lower 
limits of the reaction rate.  

Finally, the contribution to the reaction rate from
resonances above E$_{lab}$=1993 keV was computed by extrapolating our 
experimental rate to higher temperatures by following the energy dependence 
of the Hauser-Feshbach rate MOST \cite{Goriely:1998}. We did not perform 
an R-matrix analysis due to the current uncertainty of spins and 
parities of excited states in $^{23}$Na in this energy region, for which 
several experimental works have been published (for 
example see \cite{Cseh:1984,Vanderzwan:1977,Schier:1976,Kuperus:1965} ). Over 
one hundred resonances have been identified for 2.0 $\leq$ E$_{lab}$ (MeV) $\leq$ 4.7, but the 
data has not been able to constrain the spins and parities of most of the
states. The average level density is ~0.04 keV$^{-1}$, high enough to apply 
the Hauser-Feshbach formalism for calculating the reaction rate \cite{Rauscher:1997}.
At the upper limit of the Gamow window corresponding to our experimental 
data (T=1$\times$10$^9$ K), the agreement between the Hauser-Feshbach and our 
R-matrix calculated rate is very good. We used the experimentally determined 
reaction rate here to renormalize the MOST values above this temperature.

The total reaction rate consists, for temperatures below
T=1$\times$10$^9$ K, of the sum of the rate from the R-matrix 
analysis of our experimental data and the rate calculated from
extrapolated Monte Carlo cross sections at the lowest energies,
and of the Hauser-Feshbach renormalized rate above T=1$\times$10$^9$ K. 
The resulting reaction rate is shown in table \ref{theRate}.

\begin{table}
\caption{\label{tbl:F19}Reaction rate for $^{19}{F}(\alpha,p)^{22}{Ne}$ }
\begin{ruledtabular}
\begin{tabular}{lccc}
$T_9$ & $N_A<\sigma v>$ recomm & $N_A<\sigma v>$ low  & $N_A<\sigma v>$  up \\
      & $[cm^3/s$ $mol]$ & $[cm^3/s$ $mol]$ & $[cm^3/s$ $mol]$ \\
\hline
0.10 &   2.402E-22   &    1.049E-23   &    5.500E-21	\\
0.11 &   5.072E-21   &    4.173E-22   &    6.166E-20	\\
0.12 &   6.322E-20   &    8.649E-21   &    4.621E-19	\\
0.13 &   5.625E-19   &    1.158E-19   &    2.732E-18	\\
0.14 &   3.943E-18   &    1.117E-18   &    1.392E-17	\\
0.15 &   2.477E-17   &    8.120E-18   &    7.555E-17	\\
0.16 &   1.399E-16   &    4.807E-17   &    4.070E-16	\\
0.18 &   2.758E-15   &    1.073E-15   &    7.091E-15	\\
0.20 &   3.310E-14   &    1.455E-14   &    7.556E-14	\\
0.25 &   3.680E-12   &    2.047E-12   &    7.088E-12	\\
0.30 &   1.272E-10   &    7.709E-11   &    2.759E-10	\\
0.35 &   2.431E-09   &    1.484E-09   &    6.038E-09	\\
0.40 &   3.340E-08   &    1.939E-08   &    7.961E-08	\\
0.45 &   3.631E-07   &    2.064E-07   &    7.438E-07	\\
0.50 &   3.072E-06   &    1.796E-06   &    5.475E-06	\\
0.60 &   1.020E-04   &    6.176E-05   &    1.601E-04	\\
0.70 &   1.445E-03   &    8.658E-04   &    2.151E-03	\\
0.80 &   1.115E-02   &    6.630E-03   &    1.624E-02	\\
0.90 &   5.615E-02   &    3.346E-02   &    8.186E-02	\\
1.00 &   4.173E-01   &    2.483E-01   &    6.068E-01	\\
1.25 &   5.748E+00   &    3.398E+00   &    8.746E+00	\\
1.50 &   3.946E+01   &    2.278E+01   &    6.111E+01	\\
1.75 &   1.770E+02   &    1.007E+02   &    2.738E+02	\\
2.00 &   5.944E+02   &    3.381E+02   &    9.115E+02	\\
2.50 &   3.773E+03   &    2.169E+03   &    5.674E+03	\\
3.00 &   1.456E+04   &    8.458E+03   &    2.160E+04	\\
3.50 &   4.089E+04   &    2.394E+04   &    6.023E+04	\\
4.00 &   9.261E+04   &    5.448E+04   &    1.359E+05	\\
5.00 &   3.123E+05   &    1.846E+05   &    4.563E+05	\\
6.00 &   7.420E+05   &    4.398E+05   &    1.082E+06	\\
7.00 &   1.427E+06   &    8.467E+05   &    2.078E+06	\\
8.00 &   2.393E+06   &    1.421E+06   &    3.484E+06	\\
9.00 &   3.661E+06   &    2.175E+06   &    5.328E+06	\\
10.00&   5.253E+06   &    3.122E+06   &    7.643E+06	\\
\end{tabular}
\end{ruledtabular}
\label{theRate}
\end{table}

\begin{widetext}
\begin{figure}
\includegraphics[width=18cm]{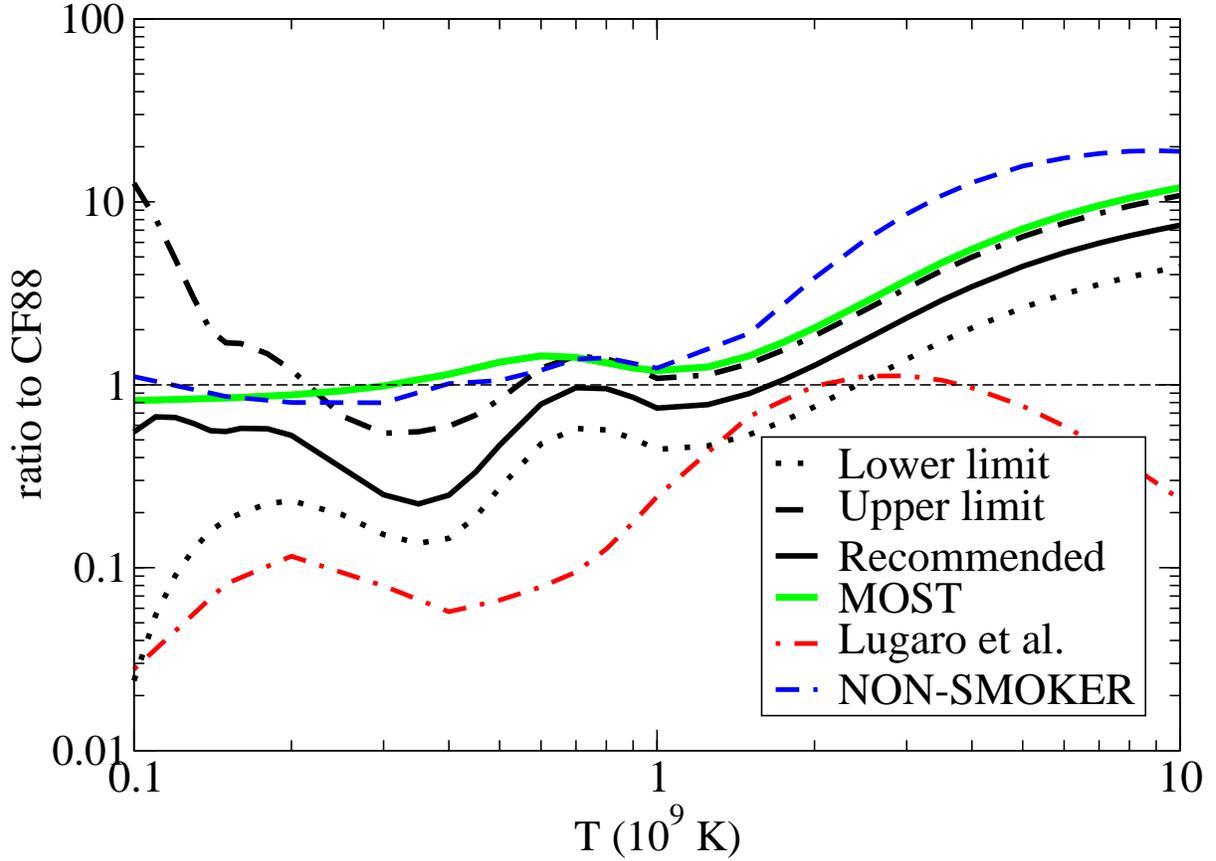}
\caption[The new reaction rate for $^{19}$F($\alpha$,p)$^{22}$Ne]
{(Color figure) The new rate for the $^{19}$F($\alpha$,p)$^{22}$Ne reaction compared
to the phenomenological rate listed in the literature (CF88) \cite{Caughlan:1988},
the Hauser-Feshbach predictions, and Lugaro {\it et al.}'s \cite{Lugaro:2004} rate.}
\label{theNewRate}
\end{figure}
\end{widetext}

Figure \ref{theNewRate} shows the total reaction rate relative to
the phenomenological rate estimate of CF88. Shown are the 
Hauser-Feshbach model predictions using the codes
MOST and NON-SMOKER relative to the CF88 predictions.
Also compared is the rate estimate of 
Lugaro {\it et al.} \cite{Lugaro:2004}, which is based on a single
non-interfering resonance approximation. 

The new rate is significantly higher (about one order of
magnitude in the stellar temperature regime) than the rate based on 
the assumption of single
non-interfering resonance levels of Lugaro {\it et al.} \cite{Lugaro:2004}. 
This can be attributed to the fact that the new rate is calculated from
non-narrow resonance contributions, as given by the R-matrix
analysis. Also, the CF88 rate is very similar to
the new rate, except for the astrophysically relevant temperature
T = 3$\times$10$^8$ K, where it is close to a factor of 10 smaller. 
However, the statistical model predictions overestimate the reaction rate 
and can be rejected at a 95\% confidence level for 2.5 $<$ T (10$^8$ K) $<$ 5.5.
For the highest temperatures, the new recommended rate differs from the statistical model
predictions by a renormalization factor (0.62). 

The main source of uncertainty in the new rate at AGB star temperatures comes from
the uncertainty in the extrapolation of the reaction cross section. Experimental
work below E$_{lab}$=800 keV is required to constrain the partial $\alpha$-particle 
widths and the interference patterns in the $^{19}$F($\alpha$,p)$^{22}$Ne reaction.
We have shown the importance of the interference effects between resonances. Therefore,
a direct measurement towards lower energies is probably the only plausible solution 
to the problem of the uncertainty of this reaction rate at AGB star temperatures.
 
For explosive stellar scenarios (T $>$ 1$\times$10$^9$ K ) the situation is still 
more delicate as spins and parities of resonances contributing to the rate are uncertain.
Both direct and indirect measurements of the $^{19}$F($\alpha$,p)$^{22}$Ne reaction above 
E$_{lab}$=2 MeV can help improving the quality of the rate for explosive scenarios.

\section{Summary and conclusions}
We have measured the $^{19}$F($\alpha$,p)$^{22}$Ne reaction in the
energy range $E_{lab}$=792-1993 keV. Stable fluorine targets were
developed and several resonances were found in the 20 experimental
yield curves. Ten different angles ranging from 30$^o$ to 150$^o$
were measured and two reaction channels ($p_0$ and $p_1$)
observed. An R-matrix analysis of the data was performed to
determine the differential and total reaction cross sections in
the investigated excitation energy range. The cross section is
characterized by many broad resonances tailing into the low
energy range. Possible additional resonance contributions in that
excitation range were predicted in a Monte Carlo cross section
analysis on the basis of available data on the nuclear level
structure of the $^{23}$Na compound nucleus. The predicted
contributions were included in the final reaction rate analysis. A
full analysis of the impact of this new rate on the fluorine
production in AGB stars will be presented in a subsequent paper.

\bibliography{f19ap}

\end{document}